\documentstyle[epsfig,multicol,prb,aps]{revtex}

\begin{document}
\title{Where are holes in Y$_{1-x}$Pr$_{x}$Ba$_{2}$Cu$_{3}$O$_{7}$?}
\author{I.I. Mazin$^{1,2}$ and A.I. Liechtenstein$^{3}$}
\address{$^1$Code 6691, Naval Research Laboratory, Washington, DC 20375\\
$^2$CSI, George Mason University\\
$^{3}$Forschungszentrum J\"{u}lich, J\"{u}lich, Germany}
\date{\today}
\maketitle

\begin{abstract}
Recent experiments by Mertz {\it et al} (Phys. Rev. {\bf B55}, 9160, 1997)
demonstrated that the loss of superconductivity upon Pr doping is associated
with the change of the character of the oxygen holes from $p_{\sigma }$ to $%
p_{\pi }.$ This experiment sheds new light onto the long-standing problem of 
$T_{c}$ suppression by Pr, and helps to rule out a number of theoretical
models,  leaving only those which predict such a transfer of the O holes.
To distinguish between the two models that do predict such an effect, one
has to access the ratio of the planar and axial character of the holes. We do
so in this paper.
\end{abstract}

\begin{multicols}{2}
One of the most exciting cases of superconductivity suppression in High-$%
T_{c}$ cuprates is that of $RE_{1-x}$Pr$_{x}$Ba$_{2}$Cu$_{3}$O$_{7},$ where $%
RE$ stands for a rare earth. A number of explanations, including magnetic
pair breaking\cite{12,27}, disorder on Ba site\cite{40}, and hole transfer
from planes to chains\cite{38}, have been suggested.
All these models have problems of various
degrees explaining the available experimental data. Most notably, recent
polarization-dependent O $1s$ near-edge X-ray absorption measurements on
single-domain Pr-doped single crystals \cite{KFK} succeeded in measuring not
only the total amount of O holes in CuO$_{2}$ planes, but also the relative
weight of their planar ($p_{x},$ $p_{y})$ and axial ($p_{z}$) components. It
is hard to overestimate the importance of this experiment, which essentially
rules out the models that assume the total hole concentration in planes to be
dependent on the Pr doping.

To the contrary, this result strongly favors the models that assume that
hole depletion of the superconducting $pd\sigma $ states occurs because of
the hole transfer into the $pd\pi $ states. The first model of this sort is
that of Fehrenbacher and Rice\cite{FR}. They assumed that $p_{\pi }$
orbitals of the planar oxygens neighboring a Pr ion, form a hybride state
with Pr $f_{z(x^{2}-y^{2})}$ orbital. They noticed that oxygens around a
rare earth site form a nearly perfect (better that 10\%) cube, so the lobes
of the $f_{z(x^{2}-y^{2})}$ at a rare earth site point nearly exactly
towards the oxygens. They assumed further that such a hybride orbital is
strongly localized. This assumption followed from the implicitly implied in
Ref. \onlinecite{FR} statement that the corresponding combination of the oxygen
orbitals forms {\it only }when the rare earth site is occupied by Pr, but
not by Y (where the $f_{z(x^{2}-y^{2})}$ state
is unoccupied and lies much higher
the O $p$ states), in other words, that direct hopping between involved O
orbitals, without assistance of the Pr $f_{z(x^{2}-y^{2})}$ orbital, is
negligible. This seemed to be a plausible idea, because in the considered O$%
_{8}$ cube the O-O distance is relatively large, $d_{O-O}\approx a/\sqrt{2}%
\approx 3$ \AA . Besides, the hopping matrix elements between two orbitals
pointing to the center of the cube is $t_{O-O}=\frac{1}{3}t_{pp\sigma }+%
\frac{2}{3}t_{pp\pi },$ more $pp\pi $ than $pp\sigma .$

On the other hand, closer look at the band structure shows that the O-O
hopping may be not small even in the absence of Pr. For instance, Andersen $%
et$ $al$\cite{Ole} found substantial $pp\sigma $ hopping perpendicular to
the planes (that is, between $p_{z}$ orbitals), of the order of 0.35 eV. O-O
hopping inside the planes is even larger, because it is assisted by the
diffuse Cu $s$ orbital (see Ref. \onlinecite{Ole} for the details). Thus, it is 
{\it a priori} unclear whether the $p_{\pi }$ orbitals should form a
localized state or rather a band comparable in width
with the superconducting $%
pp\sigma $ band. To answer this question, Liechtenstein and Mazin\cite{we}
identified among the manifold of occupied O-derived bands the oxygen band
that has the $z(x^2-y^2)$ symmetry, and found it to be rather dispersive (band
width $\approx 1.5$ eV). Based on this finding, they suggested a
modification of the original Fehrenbacher-Rice model, in which holes are
transferred from the $pd\sigma $ band not to a localized state, but to the
band derived from the Fehrenbacher-Rice states with substantial overlap
between them. Parameters of the model were found from LDA+U$_{{\rm \Pr }}$
calculations, which should be valid at small and moderate Pr doping (at
large doping, that is, close to the metal-insulator transition, one has to
take into account Hubbard U at the Cu site, which was not done in Ref.
\onlinecite
{we}). This modified Fehrenbacher-Rice model was very successful in
describing $T_{c}$ suppression rates in $RE_{1-x}$Pr$_{x}$Ba$_{2}$Cu$_{3}$O$%
_{7},$ including the dependence of the suppression rate on the rare earth.
The latter constitutes a serious problem in the original model\cite{FR} ,
because the only effect of substituting on $RE$ by another in that model is
``chemical pressure'', that is, contraction of the Cu--O and $RE$--O bonds.
The two contractions shift the energy balance in favor of the $pd\sigma $
band and the $pp\pi -f_{z(x^{2}-y^{2})}$ state, respectively\cite{FR1}.
However, the latter effect is stronger ({\it cf.} canonical scaling of
Harrison, which implies the $d^{-4}$ and $d^{-5}$ scaling for $pd$ and $pf$
hopping amplitudes, respectively). Correspondingly, external pressure
enhance $T_{c}$ suppression, the most striking manifestation being the
recently observed suppression of superconductivity by external pressure in
pure NdBa$_{2}$Cu$_{3}$O$_{6.7}$\cite{Nd}. To the contrary, bond
contractions due to host $RE$ substitutions {\it decreases} the $T_{c}$
suppression rate, opposite to what one would expect in the localized 
model of  Ref. \onlinecite{FR}. On the other hand,
in the modified model of Ref. \onlinecite{we} this effect
appears naturally as a result of hopping between the host $RE$ and the Pr $f$
orbitals.

However, closer to the metal-insulator transition the model of Ref.
\onlinecite{we}
becomes less and less realistic and it was pointed out\cite{FR1} that
uncritical extention of this model to pure PrBa$_{2}$Cu$_{3}$O$_{7}$ leads to
serious problems with explaining away possible metallic conductivity in the $%
pp\pi $ band. Thus, it is highly desirable to address the differences
between models of Refs. \onlinecite{we} and \onlinecite{FR}, by an independent
experiment. This was one of the goals of Mertz {\it et al}\cite{KFK}.
Essentially, they noticed that the $p_{\pi }$ holes have different character
in the two models: According to Ref. \onlinecite{we}, they are predominantly
planar ($p_{x,y}),$ while according to Ref.\onlinecite{FR}
the $p_{\pi }$ orbitals
in question point towards neighboring Pr, and correspondingly have
comparable amount of the $p_{x,y}$ and of $p_{z}$ character.

 As mentioned above, around each Pr eight nearest neighbor 
O ions form nearly perfect cube; correspondingly, $p$ orbitals pointing
to the Pr according to the model of Ref.\onlinecite{FR} form an angle
$\alpha=\arctan (1/\sqrt{2})\approx 35^{\circ }$ with the $xy$ plane
(in reality, the interplanar
O-O distance is slightly larger than intraplanar one, so the angle is
$\approx 36^{\circ }$ for O2 and $\approx 37^{\circ }$ for O3).
Incidentally, in the original paper\cite{FR} this angle was mistakenly
identified as 45$^{\circ} $, which did not influence the calculations of 
Fehrenbacher and Rice,\cite{FR}, but misled
the authors of Ref. \onlinecite{KFK} and
affected their comparison of the experimental results
with the Fehrenbacher-Rice model. Furthermore, as it was mentioned in Ref. 
\onlinecite{we}, in the band model the $pp\pi $ hole states are strictly planar
only in the limit when the number of $pp\pi $ holes tends to zero, {\it i.e.,%
} when the $pp\pi $ band touches the Fermi level. The experimental results
of Mertz {\it et al}\cite{KFK} correspond to substantial Pr doping ($x=0.2$)
and considerable number of the $pp\pi $ holes (0.25 holes/cell). In an
elegant analysis of their experimental data, Mertz {\it et al} deduce that
the angle that unoccupied $p_{\pi }$ orbitals form with the $xy$ plane is 20$%
^{\circ }$--25$^{\circ }$, and conclude that this is equally far from the
Fehrenbacher-Rice prediction of 45$^{\circ }$ and of the Liechtenstein-Mazin
prediction of 0$^{\circ }.$ However, as we mentioned above, the former model
predicts 35--37$^{\circ }$ instead, and the latter predicts an angle which has
yet to be calculated. Below we present such a calculation.

The original paper \cite{we} was aimed at calculating $T_{c}$ suppression
rate at small Pr concentrations, and utilized a tight-binding model where $%
p_{z}$ orbitals were not included. In order to analyze the hole character at
finite dopings, we have to include both $p_{x,y}$ and $p_{z}$ states. We
will, however, neglect the $pf\pi $ hopping and leave only the $pf\sigma $
one. The hopping amplitude between the $f_{z(x^{2}-y^{2})}$ and the $p_{x,y}$
orbitals is $\frac{\sqrt{5}}{3}\sqrt{\frac{2}{3}}t_{pf\sigma }=\sqrt{\frac{2%
}{3}}t,$ and between the $f_{z(x^{2}-y^{2})}$ and the $p_{z}$ it is $\frac{%
\sqrt{5}}{3}\sqrt{\frac{1}{3}}t_{pf\sigma }=\sqrt{\frac{1}{3}}t$ (Ref.
\onlinecite
{SKf}). Correspondingly, instead of the tight-binding Hamiltonian given in
Ref. \cite{we}, Eq. (1), we have 
\end{multicols}
\rule[10pt]{0.45\columnwidth}{.1pt}
\begin{equation}
H=\left( 
\begin{array}{ccccc}
\epsilon _{p} & 0 & 0 & 0 & t\sqrt{1/3}C_{x} \\ 
0 & \epsilon _{p} & 0 & 0 & -t\sqrt{1/3}C_{y} \\ 
0 & 0 & \epsilon _{p} & -\tau S_{x}S_{y} & t\sqrt{2/3}S_{x} \\ 
0 & 0 & -\tau S_{x}S_{y} & \epsilon _{p} & -t\sqrt{2/3}S_{y} \\ 
t\sqrt{1/3}C_{x} & -t\sqrt{1/3}C_{y} & t\sqrt{2/3}S_{x} & -t\sqrt{2/3}S_{y}
& \epsilon _{f}
\end{array}
\right) ,  \label{H}
\end{equation}
where the first two orbitals are $p_{z}$ and the remaining part is the $%
p_{x}-p_{y}-f$ Hamiltonian used in Ref. \onlinecite{we}. The same notations as
there are used, namely $S_{x,y}=2\sin (ak_{x,y}/2),$ $C_{x,y}=2\cos
(ak_{x,y}/2).$

In Ref. \onlinecite{we} we
were interested in calculating the doping dependence of
the number of unoccupied states, that is why we had to keep the $f$ orbitals
explicitly in the tight-binding Hamiltonian. Now we want to estimate the
average weight of the $p_{z}$ orbital compared to that of the $p_{x,y}$
orbitals for a given number of holes. Thus, we fold down the $f$ state in
Eq.(\ref{H}), using L\"{o}wdin perturbation theory. This gives 
\begin{equation}
H\approx \left( 
\begin{array}{c}
\begin{array}{cccc}
C_{x}^{2}\tilde{t} & \ \ -C_{x}C_{y}\tilde{t} & \ \sqrt{2}C_{x}S_{x}\tilde{t}
& \ -\sqrt{2}C_{x}S_{y}\tilde{t} \\ 
\ \ -C_{x}C_{y}\tilde{t} & \ C_{y}^{2}\tilde{t} & \ -\ \sqrt{2}C_{y}S_{x}%
\tilde{t} & \ \ 2S_{x}^{2}\tilde{t} \\ 
\sqrt{2}C_{x}S_{x}\tilde{t} & -\ \sqrt{2}C_{y}S_{x}\tilde{t} & \ 2S_{x}^{2}%
\tilde{t} & -\tau S_{x}S_{y}-2S_{x}S_{y}\tilde{t} \\ 
\ -\sqrt{2}C_{x}S_{y}\tilde{t} & \ \ \sqrt{2}C_{y}S_{y}\tilde{t} & -\tau
S_{x}S_{y}-2S_{x}S_{y}\tilde{t} & \ \ 2S_{y}^{2}\tilde{t}
\end{array}
\end{array}
\right) ,  \label{H1}
\end{equation}
\begin{flushright} \rule{0.45\columnwidth}{.1pt} \end{flushright}
\begin{multicols}{2}
where we put $\epsilon _{p}$ to zero and denote $\tilde{t}%
=t^{2}/3(E_{F}-\epsilon _{f}).$ We diagonalized this Hamiltonian numerically
and calculated the average angle for unoccupied (hole) states in the upper $%
p_{\pi }$ band as a function of filling (Fig.1) ($\alpha =\arctan \frac{%
\langle n_{x}+n_{y}\rangle }{\langle n_{z}\rangle },$ where $n_{x}+n_{y}$
and $n_{z}$ are calculated by adding squares of the first two and the last
two components of the eigenvectors). For the hole count 0.25, corresponding
to the experiment of
Ref. \onlinecite{KFK}, we find $\alpha =18^{\circ }$. Fig.1
was calculated with $\tau =0;$ for fixed hole count, the angle $\alpha $
depends on $\tau $ very little. For instance, for $\tau =\tilde{t}$ we find $%
\alpha $ to be reduced by about 20\%, that is, to 15$^{\circ }$. The latter
number can be taken as the lower bound for the band model prediction.
In fact, the reason that the band model yields an unexpectingly large
rotation angle is  that although this angle tends to zero for infinitesimally
small doping, it increases very rapidly, as the square root of hole number
and, consequently, as the square root of the Pr concentration, the fact
that was not appreciated in Ref. \onlinecite{we}.

To conclude, we have calculated the relative character of the planar, $%
p_{x,y},$ and of the axial, $p_{z},$ hole states in the $pp\pi $ band in Y$%
_{1-x}$Pr$_{x}$Ba$_{2}$Cu$_{3}$O$_{7}$ as a function of hole count in this
band, using the localized model of Fehrenbacher-Rice \cite{FR} and the band
model of Liechtenstein-Mazin\cite{we}. Comparing our results with the X-ray
absorption experiment of Mertz {\it et al}\cite{KFK}, we observe that, in
agreement with the conclusion of the authors of Ref. \onlinecite{KFK}, their
result falls {\it between} the two theoretical predictions. However, the two
predictions are closer to each other than what was supposed in Ref.
\onlinecite{KFK},
and correspondingly closer to the experiment. Especially close to the
experiment (10--30\%) are predictions of the band model of Ref.\onlinecite{we}. 

The last remark we would like to make is that, although the {\it
quantitative} difference between the two model appears to be not that
large in this case, there is a {\it qualitative }difference, which can be
used in the future experimental studies to distinguish between them. Namely, the
band model predicts that rotation angle increases with Pr doping, especially
at low doping levels, while according to the localized model it should
stay essentially constant.

 We wish to thank M.S. Golden for useful discussions, and D.J. Singh for
critical reading of the manuscript. Work at NRL is supported by the 
Office of Naval Research.

\figure{
\centerline{\epsfig{file=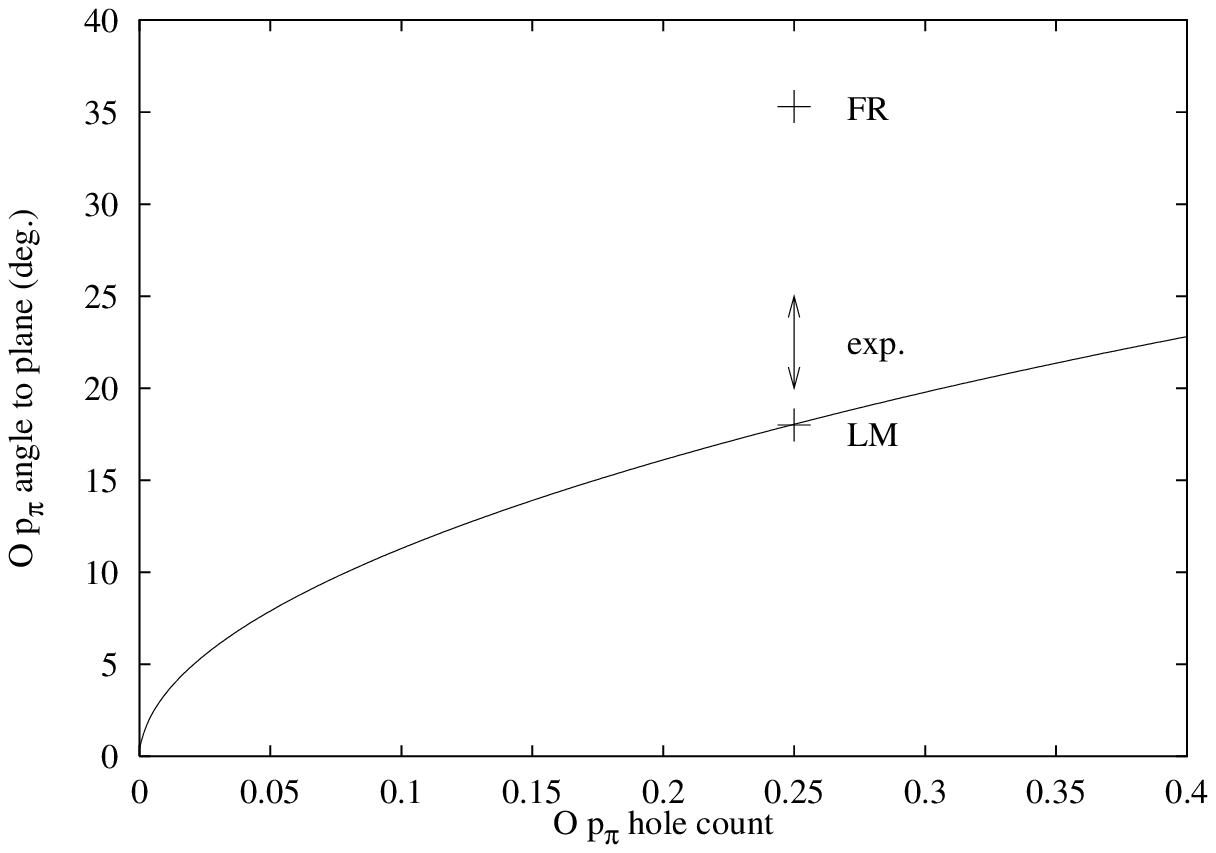,width=0.95\linewidth}}
\vspace{0.1in} \setlength{\columnwidth}{3.2in} \nopagebreak
\caption{Average rotation angle of the depleted
O(2,3) $p_\pi$ orbitals as a function of the number of holes, $n$, in
the  $pp\pi$ (``Fehrenbacher-Rice'') band, from the band model of
Liechtenstein and Mazin (LM). Experimental estimate 
at $n=0.25$ is shown by the double-headed arrow, as well as (independent
of the hole count) rotation angle in the localized Fehrenbacher-Rice
model (FR). The LM curve corresponds to $\tau=0$ in Eq.(\protect\ref{H1}). 
}}
\end{multicols}

\begin{references}
\bibitem{12}  G.Y. Guo and W. Temmerman, Phys. Rev. {\bf B 41}, 6372 (1990).

\bibitem{27}  Y. Japha and V. Zevin, Phys. Rev. {\bf B 46}, 9240 (1992).

\bibitem{40}  H.A. Blackstead and J.D. Dow, Phys. Rev. {\bf B 51}, 11 830
(1995).

\bibitem{38}  D. Khomskii, J. Super. {\bf 6}, 69 (1993).

\bibitem{KFK}  M. Mertz, N. N\"{u}cker, E. Pellegrin, P Schweiss, S.
Schuppler, M. Kielwein, M. Knupfer, M.S. Golden, J. Fink, C.T. Chen, V.
Chakarian, Y.U. Idzerda, and A. Erb, Phys. Rev. {\bf B 55}, 9160 (1997).

\bibitem{FR}  R. Fehrenbacher and T.M. Rice, Phys. Rev. Lett. {\bf 70}, 3471
(1993).

\bibitem{Ole}  O.K. Andersen, O.Jepsen, A. I. Liechtenstein, and I. I.
Mazin, Phys. Rev. {\bf B  49}, 4145 (1994).

\bibitem{we}  A.I. Liechtenstein and I.I. Mazin, Phys. Rev. Lett.,  {\bf 74}%
, 1000 (1995).

\bibitem{FR1}  R. Fehrenbacher and T.M. Rice, cond-mat/9507095.

\bibitem{Nd}  V. Dyakonov, I. Fita, N. Doroshenko, M. Baran, S. Piechota,
and H. Szymczak, Physica {\bf C 276}, 245 (1997).

\bibitem{SKf}  K. Takegahara, Y. Aoki, and A. Yanase, J. Phys.
{\bf C 13}, 583 (1980).
\end{references}
\end{document}